%% file: main.tex
\documentclass[12pt]{article}

\usepackage{graphicx}
\usepackage{siunitx}
\usepackage{hyperref}
\usepackage{amsmath, amssymb}
\usepackage{subcaption}

\title{Predicting core transport in ITER baseline discharges with neon injections}

\author{
D M Orlov$^{1}$,
J McClenaghan$^{2}$,
J Candy$^{2}$,
J D Lore$^{3}$, \\
N T Howard$^{4}$,
F Sciortino$^{4,5}$,
and C Holland$^{1}$
}

\date{}

\begin{document}
\maketitle

\begin{center}
$^{1}$Center for Energy Research, University of California San Diego, La Jolla, CA 92122, USA\\
$^{2}$General Atomics, San Diego, CA 92186, USA\\
$^{3}$Oak Ridge National Laboratory, Oak Ridge, TN 37831, USA\\
$^{4}$Massachusetts Institute of Technology, Cambridge, MA 02139, USA\\
$^{5}$Proxima Fusion, Germany\\[1ex]
Corresponding author: dorlov@ucsd.edu
\end{center}






\begin{abstract}
Achieving self-consistent performance predictions for ITER requires integrated modeling of core transport and divertor power exhaust under realistic impurity conditions. We present results from the first systematic power-flow and impurity-content study for the ITER 15 MA baseline scenario constrained directly by existing SOLPS-ITER neon-seeded divertor solutions. Using the OMFIT STEP workflow, stationary temperature and density profiles are predicted with TGYRO for $1.5 \le Z_{\rm eff} \le 2.5$, and the corresponding power crossing the separatrix $P_{\rm sep}$ is evaluated. We find that $P_{\rm sep}$ varies by more than a factor of 1.7 across this scan and matches the $\sim 100$~MW SOLPS-ITER prediction when $Z_{\rm eff} \simeq 1.6$ or when auxiliary heating is reduced to $\sim 75\%$ of nominal. Rotation-sensitivity studies show that plausible variations in toroidal flow magnitude modify $P_{\rm sep}$ by $\lesssim 20\%$, while AURORA modeling confirms that charge-exchange radiation inside the separatrix is dynamically negligible under predicted ITER neutral densities. These results identify a restricted compatibility window, $Z_{\rm eff} \approx 1.6$--1.75 and $0.75 \lesssim f_{P_{\rm aux}} \le 1.0$, in which core transport predictions remain aligned with neon-seeded divertor protection targets.  
This self-consistent, model-constrained framework provides actionable guidance for impurity control and auxiliary-heating scheduling in early ITER operation and supports future whole-device scenario optimization.
\end{abstract}



\input{Section1_Intro}          
\input{Section2_ZeffScan}       
\input{Section3_PowerFlow}      
\input{Section4_AuxPower}       
\input{Section5_Rotation}       
\input{Section7_Conclusions}    

\section*{Acknowledgments}
The authors thank the OMFIT and STEP development teams, as well as the SOLPS-ITER modeling community, for useful discussions. This work was supported by the U.S. Department of Energy, Office of Science, Office of Fusion Energy Sciences and Office of Advanced Scientific Computing Research through the Scientific Discovery through Advanced Computing (SciDAC) program under awards DE-SC0018287, DE-SC0017992, DE-AC05-00OR22725, and DE-SC001426.

\bibliographystyle{plain}

\end{document}

%% file: Section1_Intro.tex
\section{Introduction}
ITER aims to demonstrate sustained, high-performance burning plasma operation with fusion gain $Q \ge 10$ in the 15~MA H-mode baseline scenario~\cite{ITERPhysics}. Achieving this goal requires an integrated understanding of the coupled core transport, impurity dynamics, pedestal structure, and divertor heat exhaust. While substantial advances have been made in transport physics modeling \cite{yoshida_transport_2025, howard_prediction_2025, siena_first_2025}, divertor power handling \cite{kallenbach_impurity_2013, reinke_heat_2017, krieger_scrape-off_2025}, and impurity control \cite{kallenbach_seeding_2015,loarte_power_2007,putterich_determination_2019, na_integrated_2025}, obtaining fully self-consistent predictions that simultaneously satisfy both core performance and divertor protection requirements remains an open challenge for ITER scenario development.

In present-day devices, the relationship between core confinement and divertor power loads is commonly managed using impurity seeding and boundary fueling, often with neon or nitrogen injection to augment radiation in the scrape-off layer and divertor. ITER will likely rely on similar strategies to maintain divertor heat fluxes below engineering limits while preserving high fusion performance~\cite{LoarteNeSeeding}. However, increasing the impurity concentration can impact thermal confinement and affect fusion reactivity through changes in dilution, collisionality and turbulence. The required degree of impurity penetration into the confined plasma thus must be compatible with core transport predictions, particularly in the absence of ELM-driven impurity flushing.

SOLPS-ITER simulations have identified a promising baseline for ITER power-exhaust strategies under no-ELM, neon-seeded conditions~\cite{PittsSOLPSITER}. These models indicate that acceptable divertor temperatures and target heat fluxes can be achieved for power crossing the separatrix on the order of $100$~MW, provided that neon radiation in the edge remains sufficiently strong to dissipate a large fraction of the parallel heat flux. However, these SOLPS-ITER studies typically prescribe plasma profiles and $Z_{\rm eff}$ inside a narrow region within the separatrix, leaving core transport properties unconstrained by first-principles modeling.

This work addresses the critical gap between divertor-validated SOLPS-ITER modeling and first-principles core-transport predictions by delivering the first fully consistent power-flow compatibility map for ITER neon-seeded baseline operation. Unlike previous studies, we do not prescribe core profiles but predict them self-consistently using TGYRO~\cite{Candy2009}. This approach establishes the first quantitative link between impurity seeding requirements and core fusion performance constraints for ITER. Specifically, we perform predictive core transport simulations using the OMFIT STEP workflow~\cite{Meneghini2020}, which couples the EFIT free-boundary equilibrium solver~\cite{lao:1985}, the ONETWO~\cite{stjohn:1994} current-evolution model, and TGYRO~\cite{Candy2009} flux-matching transport calculations. The resulting temperature and density profiles enable direct evaluation of the core power balance and allow us to determine the target power $P_{\rm sep}$ consistent with a given impurity concentration.


Using this workflow, we quantify the sensitivity of predicted ITER power flow and performance to variations in impurity content, intrinsic rotation, and fueling sources. A baseline case is constructed using a CORSICA equilibrium for the 15~MA H-mode inductive scenario and compared directly with published SOLPS-ITER solutions for a neon-seeded divertor. By scanning the core-effective charge in the range $1.5 \le Z_{\rm eff} \le 2.5$, while holding fixed the impurity composition, separatrix conditions, and neutral fueling profiles prescribed by the SOLPS-ITER modeling, we identify a narrow region of compatibility between core transport predictions and the divertor scenario. Here, compatibility is defined as the simultaneous satisfaction of steady-state core power balance, target power exhaust constraints, and consistency between core radiation levels and the impurity and neutral sources predicted by SOLPS-ITER. We further assess the sensitivity of these results to uncertainties in the prescribed core toroidal rotation amplitude and to the influence of charge exchange using the AURORA impurity-evolution framework~\cite{SciortinoAURORA}.
The results presented here demonstrate that whole-device consistency criteria place non-trivial constraints on viable ITER operating points and highlight the need for transport-aware impurity control strategies. The methodology developed in this study is extensible to future integrated-model validation against ITER-relevant experiments on existing tokamaks and supports predict-first tools for scenario optimization in ITER.

The simulations presented here are based on the baseline CORSICA \cite{Crotinger1997_CORSICA} equilibrium and SOLPS-ITER neon-seeded divertor solutions available at the time of this work. These edge solutions were developed for neon injection scenarios, whereas ITER has since moved toward a full high-$Z$ tungsten wall environment \cite{Pitts2025_PWI_impact}. Under such metallic-wall conditions, tungsten sourcing, screening and radiation may play a role comparable to or even exceeding that of neon in determining the pedestal structure, radiated power fraction, and core impurity balance. In the present study, however, the dominant radiative and dilution effect arises from neon, and the self-consistent transport response is primarily set by $Z_{\mathrm{eff}}$ and not by tungsten concentration. A dedicated follow-on study will therefore be required to evaluate combined Ne+W operation (e.g. as in Fajardo \textit{et al}~\cite{fajardo_theory-based_2025}) and to determine whether the compatibility window identified here remains stable, shifts, or narrows when tungsten transport, accumulation, and pedestal screening physics are included in the OMFIT STEP workflow.  

The paper proceeds by first introducing the ITER baseline equilibrium and a controlled $Z_{\rm eff}$ scan used to quantify the role of impurity concentration (Section~\ref{sec:ZeffScan}). We then determine the resulting core power balance and identify cases that remain consistent with SOLPS-ITER neon-seeded divertor solutions (Section~\ref{sec:PowerFlow}). Sensitivity studies examining auxiliary heating, rotation and particle sources (Sections~\ref{sec:AuxPower}--\ref{sec:Rotation}) define a restricted compatibility window for ITER operations. The main results and implications for ITER scenario optimisation are summarised in Section~\ref{sec:Conclusions}.

%% file: Section2_ZeffScan.tex
\section{Setting up ITER profiles for a controlled $Z_{\rm eff}$ scan}
\label{sec:ZeffScan}

Predicting the impact of impurity concentration on core performance requires a consistent treatment of plasma composition, dilution and radiation. We therefore construct a controlled scan of the effective charge, $Z_{\rm eff}$, representative of neon-seeded divertor operation in ITER.


All simulations begin from a standard CORSICA equilibrium~\cite{Crotinger1997_CORSICA} representative of the ITER 15~MA full-bore inductive baseline scenario during burn, corresponding to $Q\simeq10$ operation with high-performance H-mode
confinement. The equilibrium geometry and kinetic profiles are consistent with ASTRA \cite{Pereverzev2002_ASTRA} scenario simulations for this case~\cite{Polevoi2002}, and feature a single-null divertor configuration with toroidal field $B_T=-5.3\,\mathrm{T}$, elongation $\kappa\simeq1.8$, and a broad region of flat safety factor in the core ($q_0<1$).

The pedestal density and temperature are prescribed rather than predicted self-consistently, reflecting the non-ELMing character of the reference scenario. Pedestal values, $T_{e,\mathrm{ped}}\simeq4.5\,\mathrm{keV}$ and $n_{e,\mathrm{ped}}\simeq8\times10^{19}\,\mathrm{m^{-3}}$, are taken from ASTRA modeling and correspond to a pedestal pressure close to the peeling–ballooning stability limit expected for ITER baseline operation. The core plasma reaches $T_{e0}\simeq T_{i0}\approx29\,\mathrm{keV}$ and $n_{e0}\approx8.5\times10^{19}\,\mathrm{m^{-3}}$.

Auxiliary heating consists of $33\,\mathrm{MW}$ of deuterium neutral beam injection and $18\,\mathrm{MW}$ of electron cyclotron power, consistent with the reference inductive ITER heating mix used in the ASTRA scenario calculations~\cite{Polevoi2002}. These values are held fixed in the baseline case and varied parametrically in subsequent power-scan studies.

Since first-principles predictions of intrinsic rotation in ITER remain uncertain, we prescribe a representative core toroidal rotation frequency $\Omega_{\mathrm{rot},0}=-20\,\mathrm{krad/s}$, consistent with extrapolations from present low-torque devices~\cite{Chrystal2017,Li2019}. The corresponding radial rotation profiles shown in Fig.~\ref{fig:RotationProfiles} are obtained by applying this core value as a rigid shift to a reference rotation shape, resulting in weak core gradients consistent with low external torque conditions.

\begin{figure}[t]
\centering
\includegraphics[width=0.32\textwidth]{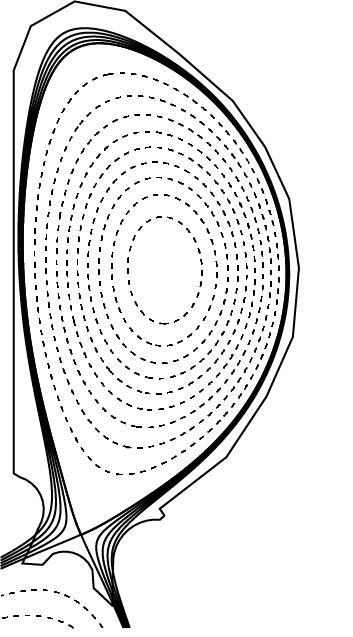}
\includegraphics[width=0.32\textwidth]{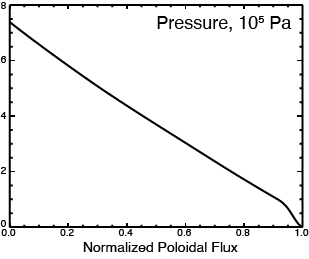}
\includegraphics[width=0.32\textwidth]{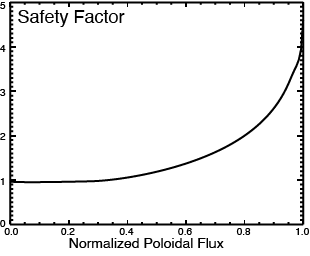}
\caption{Baseline ITER 15~MA single-null equilibrium and initial kinetic profiles used as the starting point for all simulations in this study. From left to right: magnetic geometry with the separatrix indicated; electron pressure profile; and safety-factor profile $q(r)$, characterized by a broad region of flat $q$ ($q_0 < 1$) typical of the ITER baseline scenario.}
\label{fig:ITERbaseline}
\end{figure}

The equilibrium geometry and initial kinetic profiles are shown in Fig.~\ref{fig:ITERbaseline}, and serve as the starting point for the subsequent $Z_{\rm eff}$, auxiliary power, and rotation sensitivity scans.

Initial kinetic profiles for all species are generated using the \textsc{PRO\_create} module within OMFIT, which sets pedestal constraints consistent with H-mode scaling and ensures compatibility with the EFIT equilibrium reconstruction. The profile shapes are parameterized in the same way as is done in the EPED code for generating equilibria \cite{slendebroek_elevating_2023}. Deuterium, tritium, helium and neon are included as active species, consistent with the assumptions used in the SOLPS simulation Additionally, no fast particles (either $\alpha$ fusion products or from auxiliary heating systems) are included in the analysis.  The impurity density profiles are assumed to be radially flat in order to maintain a constant core-effective charge, $Z_{\rm eff}$. This assumption is adopted to reduce the dimensionality of the parameter scan and to isolate the impact of varying $Z_{\rm eff}$ on core transport and power-flow predictions.
As discussed further below in Section \ref{sec:PowerFlow}, the SOLPS-ITER boundary conditions used in this work constrain impurity behavior primarily in the SOL and within a narrow region inside the LCFS, where the neon-seeded divertor solutions are well validated. Inside this boundary, the neon concentration is continued inward using a linear extrapolation to generate a radially flat $Z_{\mathrm{eff}}$ profile for STEP/TGYRO. While this provides a controlled way to explore core transport sensitivity to effective charge, it does not capture potential impurity screening or accumulation across the EPED-scale pedestal — effects that may arise from rotation shear, turbulent pinch/anti-pinch physics, or ELM flushing. A full core–edge integrated modeling chain, incorporating impurity transport solvers such as \textsc{STRAHL} coupled to SOLPS–ITER and STEP, will therefore be required to assess whether neon penetration modifies the compatible $(Z_{\mathrm{eff}},P_{\mathrm{aux}})$ operating window identified here. Such multi-code simulations are computationally intensive and beyond the scope of this manuscript, but represent a natural next step toward truly integrated core–edge performance prediction for ITER.

\begin{figure}
  \centering

  \begin{subfigure}{0.45\textwidth}
    \includegraphics[width=\linewidth]{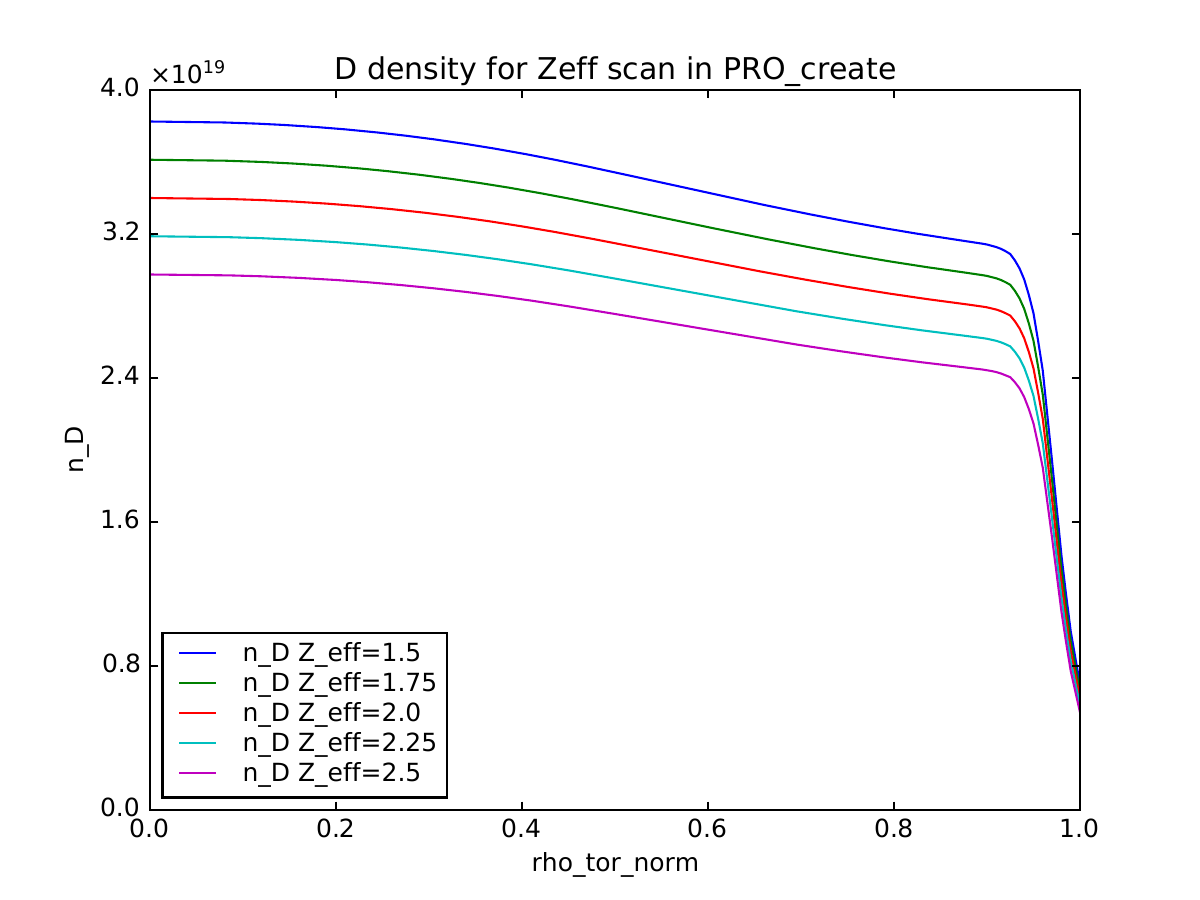}
    \caption{Deuterium}
  \end{subfigure}
  \hfill
  \begin{subfigure}{0.45\textwidth}
    \includegraphics[width=\linewidth]{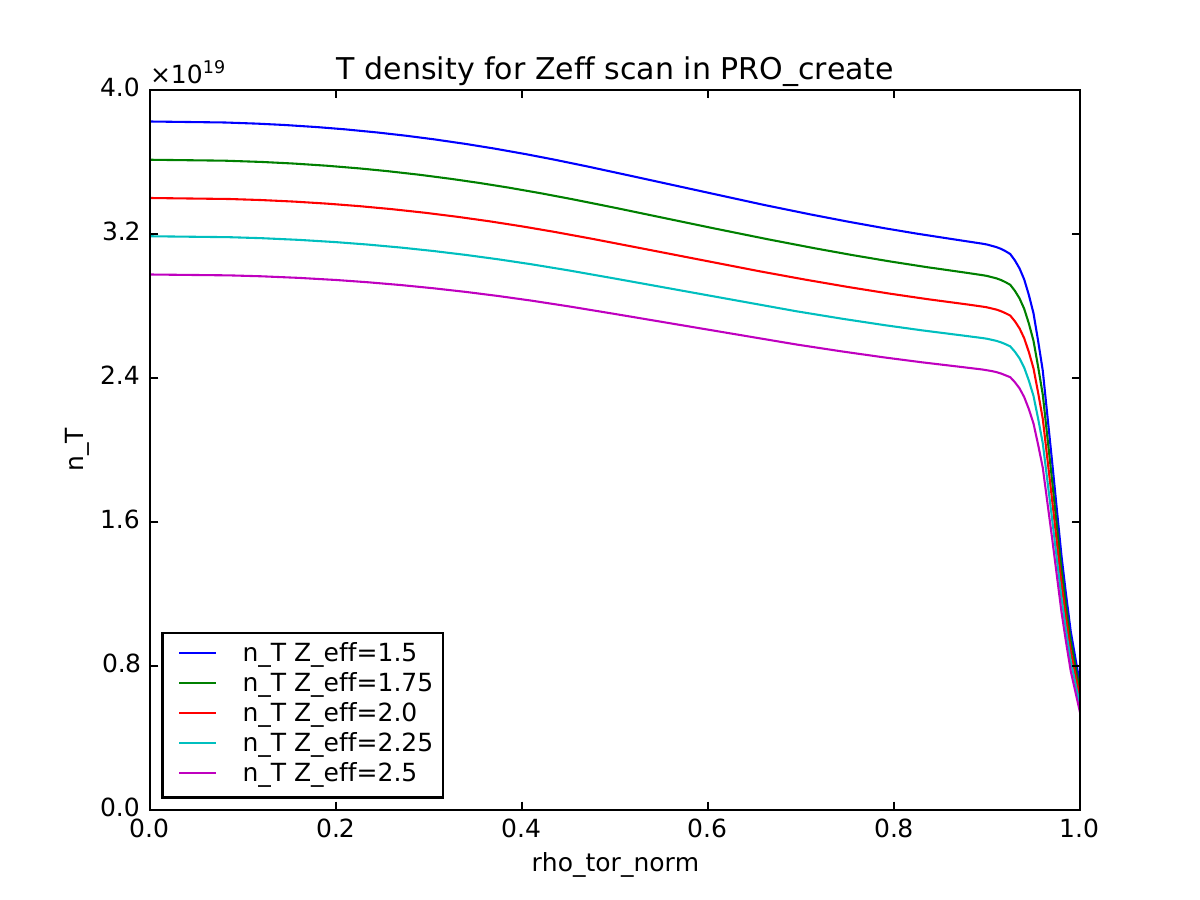}
    \caption{Tritium}
  \end{subfigure}

  \vspace{0.5em}

  \begin{subfigure}{0.45\textwidth}
    \includegraphics[width=\linewidth]{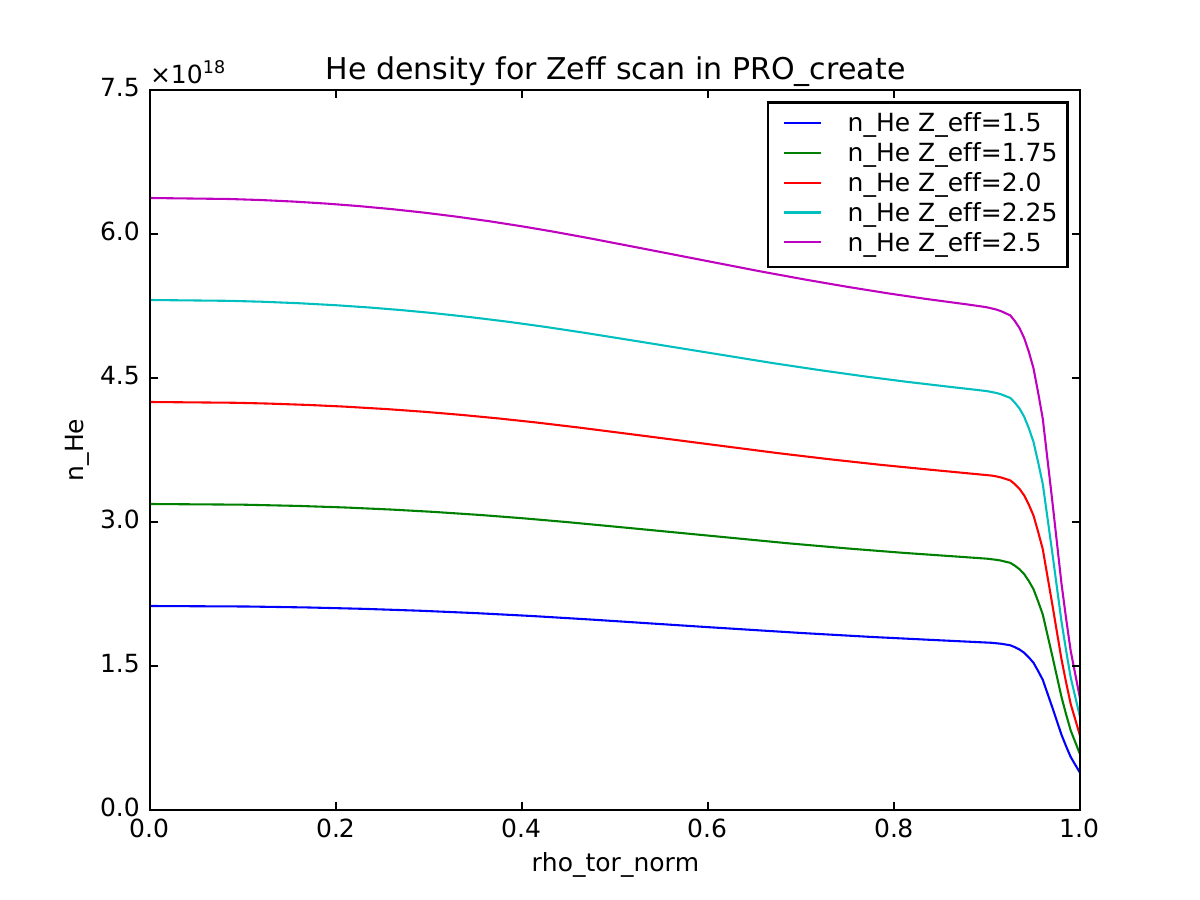}
    \caption{Helium}
  \end{subfigure}
  \hfill
  \begin{subfigure}{0.45\textwidth}
    \includegraphics[width=\linewidth]{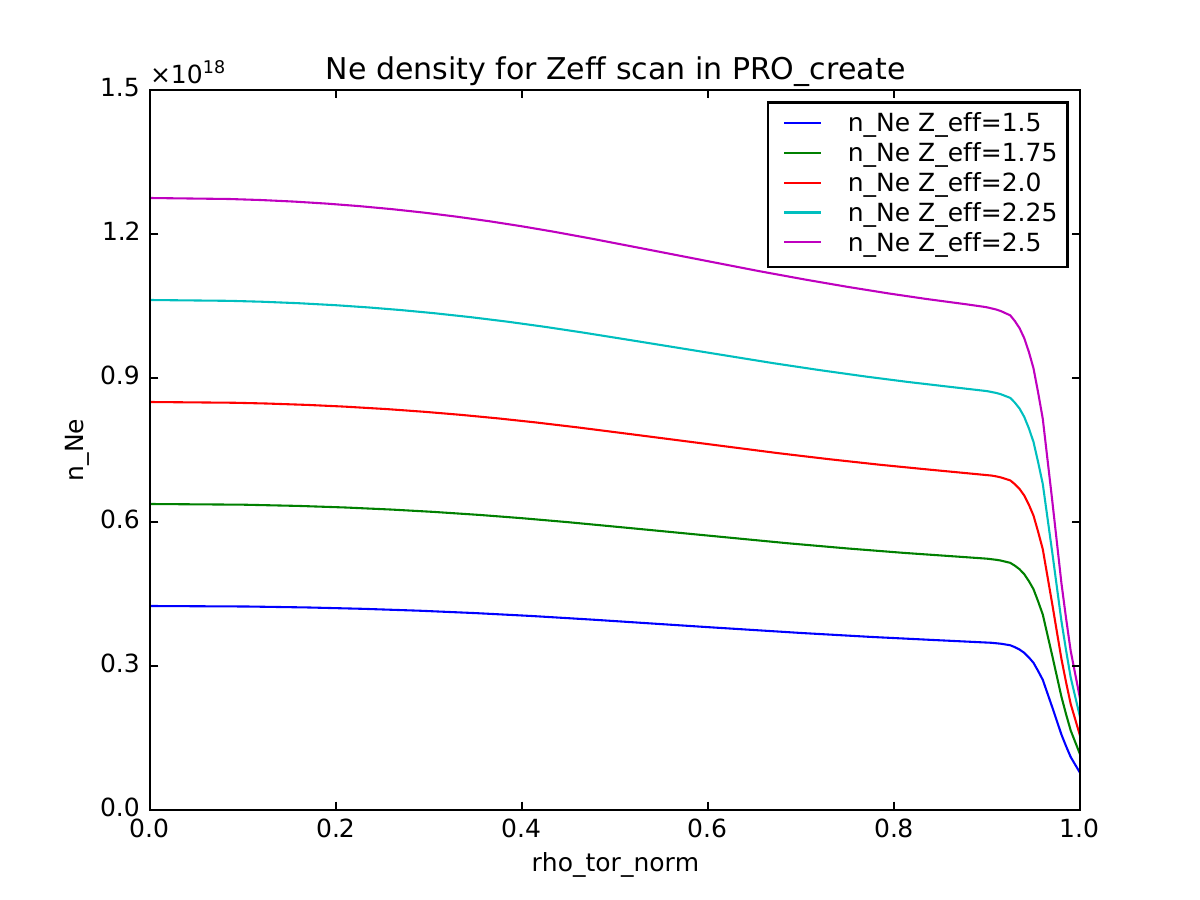}
    \caption{Neon}
  \end{subfigure}

  \caption{Initial core density profiles for deuterium, tritium, helium,
  and neon for the $Z_{\rm eff}$ scan in PRO-CREATE.}
  \label{fig:ZeffDensityScan}
\end{figure}

A five-point scan, $Z_{\rm eff} = 1.5,\,1.75,\,2.0,\,2.25,\,2.5$, is constructed to cover the range expected during neon-seeded operation consistent with acceptable divertor heat-flux dissipation in SOLPS-ITER modelling. In this scan, $Z_{\rm eff}$ is varied by adjusting the impurity content while keeping the underlying deuterium--tritium fuel composition fixed, with helium and neon impurities weighted to contribute equally to the total effective charge. Higher $Z_{\rm eff}$ values therefore correspond to increased impurity radiation but also increased core dilution, while lower values reduce radiative power dissipation and may violate divertor power-exhaust constraints. Figure~\ref{fig:ZeffDensityScan} shows the resulting initial radial density profiles for D, T, He, and Ne. The predicted helium concentration increases with $Z_{\rm eff}$ due to fusion-born He ash content assumptions in \textsc{PRO\_create}. For $Z_{\rm eff}=1.75$, the neon concentration is consistent with values that produce ${\sim}100\,\mathrm{MW}$ of power crossing the separatrix in published SOLPS-ITER simulations for no-ELM neon-seeded operation.

These kinetic profiles and the equilibrium are used as the initial conditions for subsequent STEP/TGYRO simulations in  Section~\ref{sec:PowerFlow}, which self-consistently solve for the stationary transport state and enable direct comparison to SOLPS-ITER power-exhaust predictions.

%% file: Section3_PowerFlow.tex
\section{Core power-flow predictions for varying $Z_{\rm eff}$}
\label{sec:PowerFlow}

We now determine the power transported across the separatrix in each $Z_{\rm eff}$ case introduced in Section~\ref{sec:ZeffScan} by evolving the temperature, density and rotation profiles to a stationary transport solution using the OMFIT STEP workflow. In this study, STEP is used to couple the EFIT free-boundary equilibrium solver, the ONETWO current diffusion capabilities, the EPED code, and the TGYRO transport solver to predict self-consistent kinetic profiles and magnetic equilibria. Here, TGYRO is used to iterate the local density and temperature gradients at a set of specified radii ($\rho_{tor} = \{0.3,0.4,...,0.9 \}$) until the predicted turbulent and neoclassical fluxes (calculated using the TGLF SAT\textbf{X} model and NEO code \textbf{refs} respectively) match the externally applied heating and current drive~\cite{Candy2009,Meneghini2020}. The fusion-born $\alpha$ source is included, consistent with the D--T burning operation expected for ITER, but all $\alpha$ particles are assumed to be fully thermalized with no fast population included.

\begin{figure}
\centering
\includegraphics[width=0.75\textwidth]{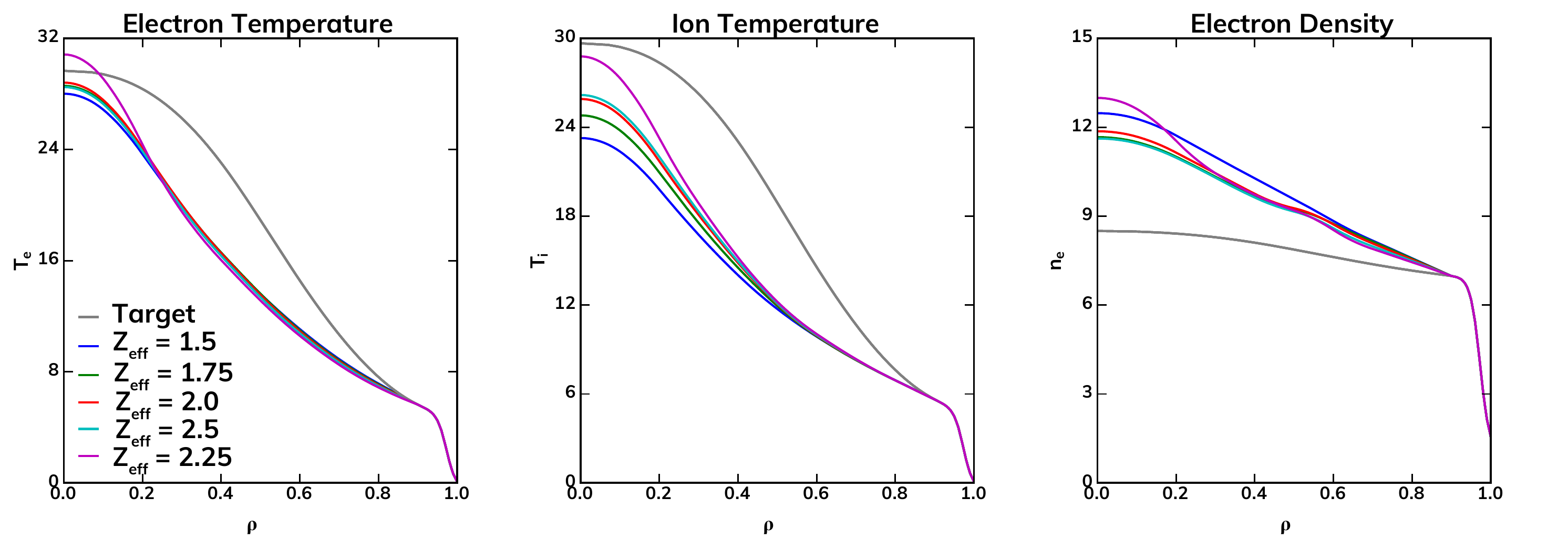}
\caption{Stationary temperature and density profiles predicted by TGYRO for the $Z_{\rm eff}$ scan. Rising $Z_{\rm eff}$ leads to increased electron collisionality and correspondingly lower core temperatures, while ion heating becomes relatively more efficient.}
\label{fig:TgyroProfiles}
\end{figure}

Figure~\ref{fig:TgyroProfiles} shows the resulting stationary profiles for electron density and for electron and ion temperatures. Compared to the initial \textsc{PRO\_create} profiles, TGYRO predicts broader density profiles and reduced core temperatures driven by the increased collisional transport associated with higher impurity concentrations. Notably, the ion temperature increases with $Z_{\rm eff}$ in the outer core, reflecting a shift toward stronger ion heating relative to electron heat conduction in highly diluted plasmas.
In these calculations, TGYRO evolves the $T_i$, $T_e$, and $n_e$ profiles such that the predicted electron particle flux matches the values expected from the separately calculated and input neutral beam fueling source. The various ion species are assumed to take the same shape as the electron density in order to maintain a constant $Z_{\rm eff}$, while all ions have the same temperature.

\begin{figure}
\centering
\includegraphics[width=0.75\textwidth]{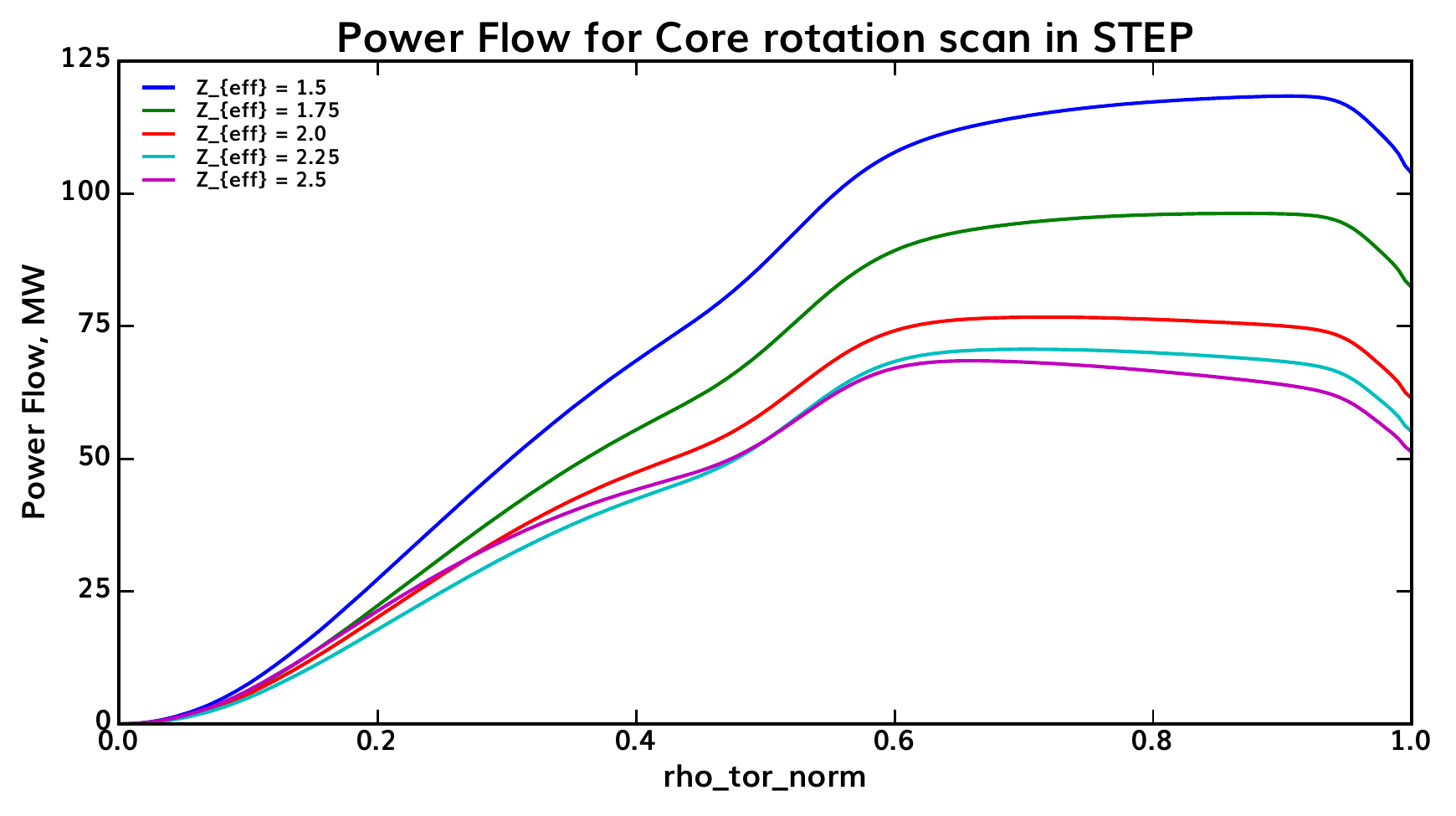}
\caption{Radial power-flow profiles predicted by TGYRO for five values of $Z_{\rm eff}$. The extracted power crossing the separatrix decreases monotonically with increasing impurity concentration, and matches the SOLPS-ITER target condition of ${\sim}100\,\mathrm{MW}$ for $Z_{\rm eff} \approx 1.75$.}
\label{fig:PowerZeffScan}
\end{figure}

The key metric for comparison with SOLPS-ITER divertor simulations is the target power (power crossing the separatrix), computed as the net outward thermal and alpha heating flux at $\rho = 0.9$. The resulting power-flow profiles are shown in Fig.~\ref{fig:PowerZeffScan}. All cases exhibit a flattening of the power flux in the outer region $0.6 \lesssim \rho \lesssim 0.9$, yielding a well-defined estimate for $P_{\rm sep}$.

The predicted power crossing the separatrix varies strongly with impurity content, ranging from ${\sim}120\,\mathrm{MW}$ at $Z_{\rm eff} = 1.5$ to ${\sim}70\,\mathrm{MW}$ at $Z_{\rm eff} = 2.5$. The $Z_{\rm eff} = 1.75$ case yields $P_{\rm sep} \simeq 100\,\mathrm{MW}$, in direct agreement with the power-flow assumed in SOLPS-ITER neon-seeding simulations for no-ELM ITER operation~\cite{PittsSOLPSITER}. This identifies a narrow compatibility window in which both core confinement and edge power-exhaust requirements can be satisfied simultaneously.

The TGYRO transport solutions obtained in this work can be compared with previous SOLPS-ITER simulations of the same neon-seeded ITER divertor scenario, which provide an independent constraint on the impurity content inside the separatrix. The SOLPS-ITER simulations are from IMAS (number 123303), which is the baseline case used for high main ion fueling and neon seeding scans as described in \cite{Lore_2022}. These simulations have $P_{SOL} = 100$ MW and cross-field diffusivities in the SOL corresponding to a heat flux width of 3.4 mm. An H-mode-like transport barrier is applied in the core by reducing the diffusivities for the first 6 mm inside of the separatrix. Cross-field drifts are not activated. 

\begin{figure}[t]
\centering
\includegraphics[width=0.35\textwidth]{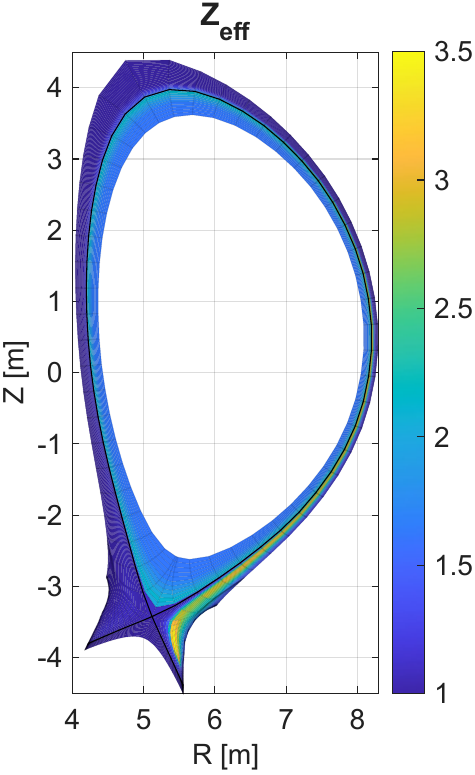}
\hspace{4mm}
\includegraphics[width=0.38\textwidth]{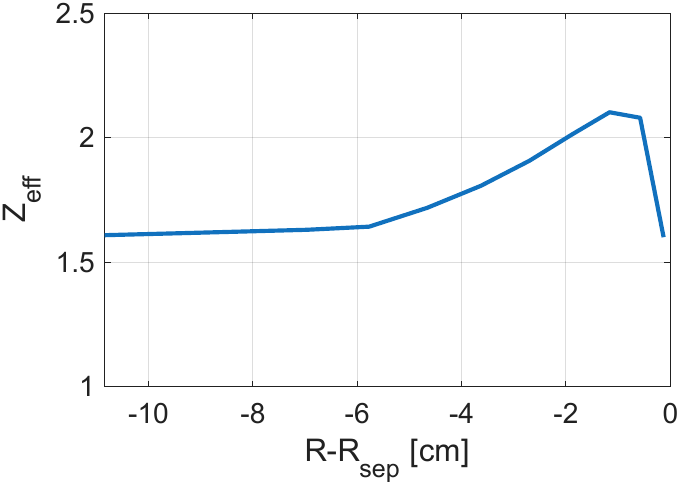}
\caption{SOLPS-ITER simulation of the neon-seeded ITER divertor scenario used to constrain the edge impurity content in this work. Left: two-dimensional distribution of $Z_{\rm eff}$ in the SOLPS-ITER domain, including the scrape-off layer and a narrow band of closed flux surfaces. While the open-field-line region exhibits strong poloidal variation in impurity concentration, the closed-field-line region is nearly uniform, justifying the use of a radially flat $Z_{\rm eff}$ in the core-transport modeling \cite{Lore_2022}. Right: poloidally averaged radial profile of the effective charge $Z_{\rm eff}$ just inside the last closed flux surface, plotted versus outer mid-plane distance from the separatrix $dR^{\rm OMP}_{\rm sep}$. The profile rapidly flattens to $Z_{\rm eff}\simeq 1.6$ in the confined region.}
\label{fig:SOLPS_Zeff}
\end{figure}

Figure~\ref{fig:SOLPS_Zeff} summarizes the relevant SOLPS-ITER results. The large panel on the right shows the two-dimensional distribution of $Z_{\rm eff}$ in the SOLPS-ITER domain, including the scrape-off layer and a narrow region of closed flux surfaces. In the open-field-line region, where recycling and wall interactions dominate, the impurity concentration is strongly non-uniform in both radial and poloidal directions. In contrast, within the confined region just inside the separatrix, the $Z_{\rm eff}$ distribution becomes nearly poloidally uniform and therefore suitable for flux-surface averaging.

The smaller panel on the lower left displays the corresponding poloidally-averaged $Z_{\rm eff}$ profile as a function of the outer mid-plane distance from the separatrix $dR^{\rm OMP}_{\rm sep}$. Just inside the last closed flux surface the profile rapidly flattens to a value $Z_{\rm eff} \simeq 1.6$, in close agreement with the baseline parameter used in our core-transport scans. Motivated by this consistency, we fix $Z_{\rm eff}=1.6$ in the OMFIT STEP/TGYRO workflow and scan the auxiliary heating power in the range $0.6 \le f_{\mathrm{Paux}} \le 1.2$ around the nominal 33~MW of NBI and 18~MW of EC. As discussed in Section~\ref{sec:AuxPower}, the resulting variation in $P_{\rm sep}$ identifies an operating interval in which the core transport predictions remain compatible with the SOLPS-ITER divertor solution.

As shown later in Section~\ref{sec:Conclusions}, this defines the core portion of a narrow operational window jointly compatible with divertor-protection and fusion-performance requirements.

These results demonstrate that even small changes in $Z_{\rm eff}$ can shift ITER power balance by more than ${\sim}50\%$. While operation at higher impurity levels eases divertor heat flux requirements, the associated reduction in fusion power and alpha heating could compromise the ITER performance mission if not compensated by auxiliary heating or improved confinement. In the following sections, we therefore investigate how $P_{\rm sep}$ responds to adjustments in auxiliary power input (Section~\ref{sec:AuxPower}), intrinsic rotation amplitude (Section~\ref{sec:Rotation}), and particle-source assumptions, which are expected to be uncertain in early ITER operation.

Overall, the $Z_{\rm eff}$ dependence obtained here establishes one of the first quantitatively validated links between impurity seeding levels and core power exhaust for ITER, providing a physics-informed basis for integrated scenario optimization.

We also assessed the potential impact of charge-exchange (CX) processes on core radiation using the AURORA impurity-evolution module coupled to the stationary TGYRO solutions. Neutral densities were taken from SOLPS-ITER modeling for the same neon-seeded baseline scenario. For all cases examined, CX radiation was found to contribute negligibly to the total core radiated power ($<1\%$), with changes in effective charge $\Delta Z_{\rm eff}<0.01$. This result reflects the extremely low neutral densities predicted inside the separatrix for ITER baseline conditions and indicates that CX processes do not affect the power-balance compatibility window identified here.

%% file: Section4_AuxPower.tex
\section{Auxiliary power scan at fixed $Z_{\rm eff}$}
\label{sec:AuxPower}

The results of Section~\ref{sec:PowerFlow} show that matching SOLPS-ITER power-exhaust predictions requires a specific range of impurity content. However, even within that compatibility window, the achievable core temperatures and fusion-power production remain sensitive to the balance between external auxiliary heating and self-heating from fusion alphas. Since the expected heating mix in ITER may vary during commissioning and early operational phases, we explore the dependence of the stationary power flow on the auxiliary heating level.

A reference case with $Z_{\rm eff} = 1.6$ is selected, consistent with the SOLPS-ITER neon-seeding scenario that predicts a power crossing the separatrix of ${\sim}100\,\mathrm{MW}$ under divertor-protection constraints~\cite{PittsSOLPSITER}. We scale the nominal $P_{\rm aux} \simeq 50 \, \rm MW$ of external heating sources 
uniformly over the range $0.6 \le f_{\mathrm{Paux}} \le 1.2$, and solve each case to stationarity with TGYRO.

\begin{figure}
\centering
\includegraphics[width=0.75\textwidth]{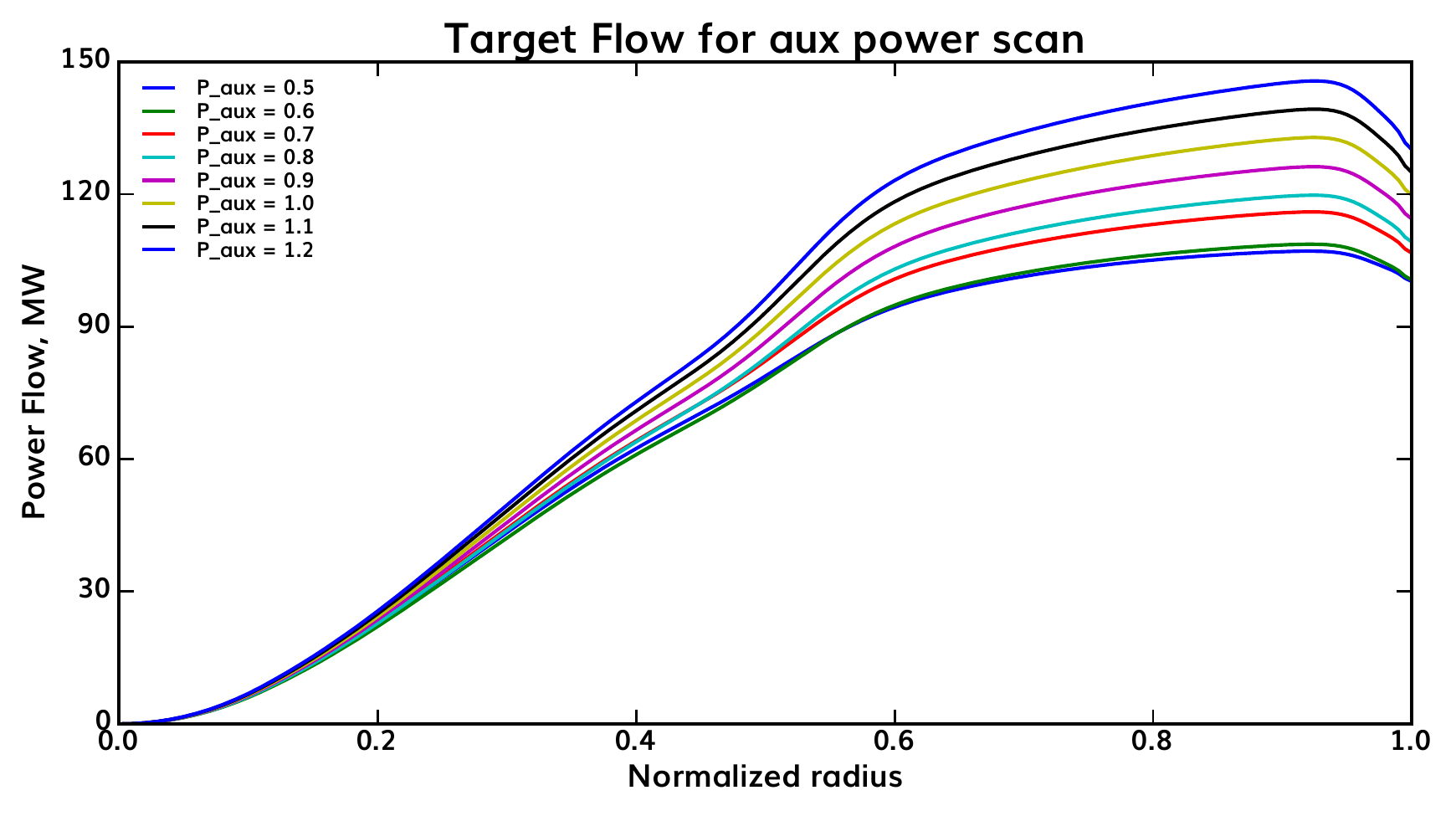}
\caption{Predicted power-flow profiles from TGYRO for a scan of auxiliary heating at fixed $Z_{\rm eff} = 1.6$. Reducing external heating lowers the power crossing the separatrix, with $f_{\mathrm{Paux}}{\sim}0.75$ producing $P_{\rm sep} \simeq 100\,\mathrm{MW}$ in agreement with SOLPS-ITER neon-seeded modelling.}
\label{fig:Zeff16PauxScan}
\end{figure}

Figure~\ref{fig:Zeff16PauxScan} shows that the predicted power crossing the separatrix increases monotonically as the auxiliary heating is increased. Crucially, the case $f_{\mathrm{Paux}} \simeq 0.75$ yields $P_{\rm sep} \simeq 100\,\mathrm{MW}$, i.e.\ matching the target condition for the SOLPS-ITER divertor solution. This solution provides a second, independent operating point in which whole-device  consistency between core transport and divertor heat-flux requirements is satisfied. For reference, the Martin 2008 scaling~\cite{martin_power_2008} predicts an L-H power threshold of $P_{L\text{--}H}\approx 75$--$85\,\mathrm{MW}$ for ITER baseline parameters (using $B_T=5.3\,\mathrm{T}$, $\bar n_e\approx0.8\times10^{20}\,\mathrm{m^{-3}}$, and $S\approx7.1\times10^2\,\mathrm{m^2}$), placing the $P_{\rm sep}\simeq100\,\mathrm{MW}$ compatibility point $\sim$20--35\% above the nominal threshold. 

This result highlights a trade-off that will require careful optimization in ITER. Lower auxiliary heating improves compatibility with divertor constraints but reduces fusion-alpha amplification, making it more difficult to reach and sustain the Q~$\ge10$ mission goals. Conversely, near-nominal heating levels increase fusion performance but may demand significantly stronger impurity radiation or alternative power-handling techniques to protect the divertor.

The $f_{\mathrm{Paux}}$ scan thus establishes that maintaining $P_{\rm sep}\simeq100\,\mathrm{MW}$ can be accomplished either by moderately reducing auxiliary power or by operating near $Z_{\rm eff}\simeq1.75$ (Section~\ref{sec:PowerFlow}). Combining these results yields a two-parameter map of operating space that constrains early ITER scenario development. In order to reduce the uncertainty in this map, we next evaluate the effects of intrinsic plasma rotation and particle-source assumptions, which introduce additional variability in transport predictions (Section~\ref{sec:Rotation}).

%% file: Section5_Rotation.tex
\section{Sensitivity to toroidal rotation amplitude and shear}
\label{sec:Rotation}

We next examine the sensitivity of the predicted power flow to variations in the prescribed toroidal rotation profile and associated rotation shear. Rather than predicting rotation self-consistently, we impose a range of toroidal rotation amplitudes representative of expected low-torque ITER operation and assess their impact on transport and power-exhaust predictions. Predicting intrinsic plasma rotation in ITER remains a significant challenge, as momentum input from neutral beams is expected to be relatively small compared to present devices~\cite{Chrystal2017,Li2019}. Uncertainties in the toroidal rotation profile can nevertheless influence turbulent transport through rotational shear stabilization and thus modify the core power balance. To quantify this effect, we perform a rotation-sensitivity scan at fixed $Z_{\rm eff}=1.6$ and nominal auxiliary heating.

\begin{figure}
\centering
\includegraphics[width=0.75\textwidth]{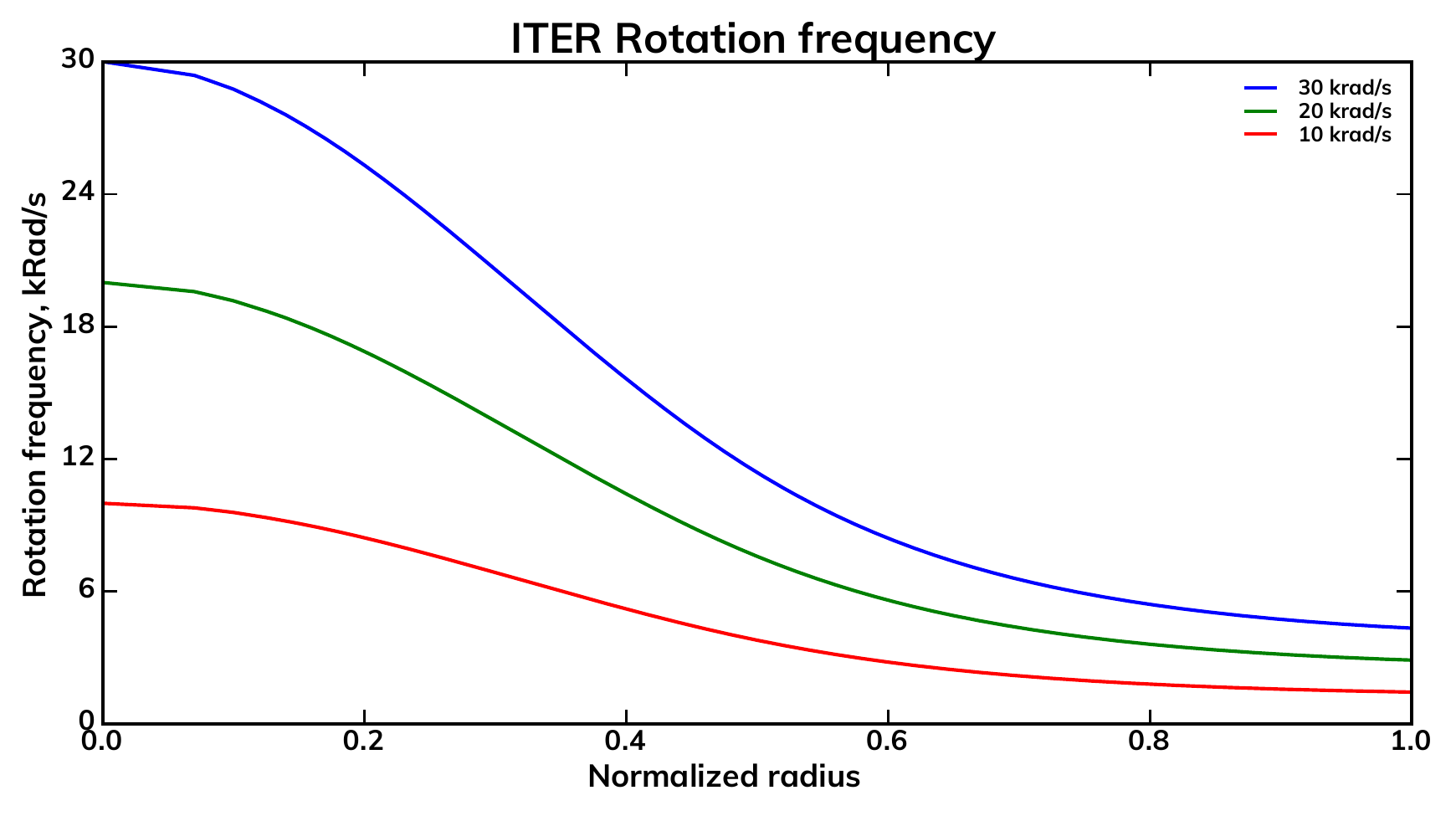}
\caption{Toroidal rotation profiles used in the rotation‐sensitivity scan at fixed $Z_{\rm eff} = 1.6$, covering the range of intrinsic rotation amplitudes expected for ITER baseline scenarios.}
\label{fig:RotationProfiles}
\end{figure}

We impose toroidal rotation profiles with core frequencies $\Omega_{\rm rot,0} = 0,\,10,\,20,$ and $30\,\mathrm{krad/s}$, applied as rigid shifts to the equilibrium profile used in Section~\ref{sec:PowerFlow}. The value $\Omega_{\rm rot,0} \simeq 20\,\mathrm{krad/s}$ is representative of the most probable intrinsic rotation estimate for the ITER baseline scenario based on extrapolations from current devices under low‐torque conditions~\cite{Chrystal2017,Li2019}. The other cases bracket the range of likely variation in early ITER operation. The corresponding rotation profiles are shown in Fig.~\ref{fig:RotationProfiles}. 
Across all cases examined, the imposed toroidal rotation shear remains modest, with $|R\,\partial\Omega_\phi/\partial r| \lesssim (3$--$5)\times10^4~\mathrm{s^{-1}}$, consistent with expectations for low-torque ITER operation where external momentum input is minimal \cite{rice_intrinsic_rotation_2007}.

\begin{figure}
\centering
\includegraphics[width=0.75\textwidth]{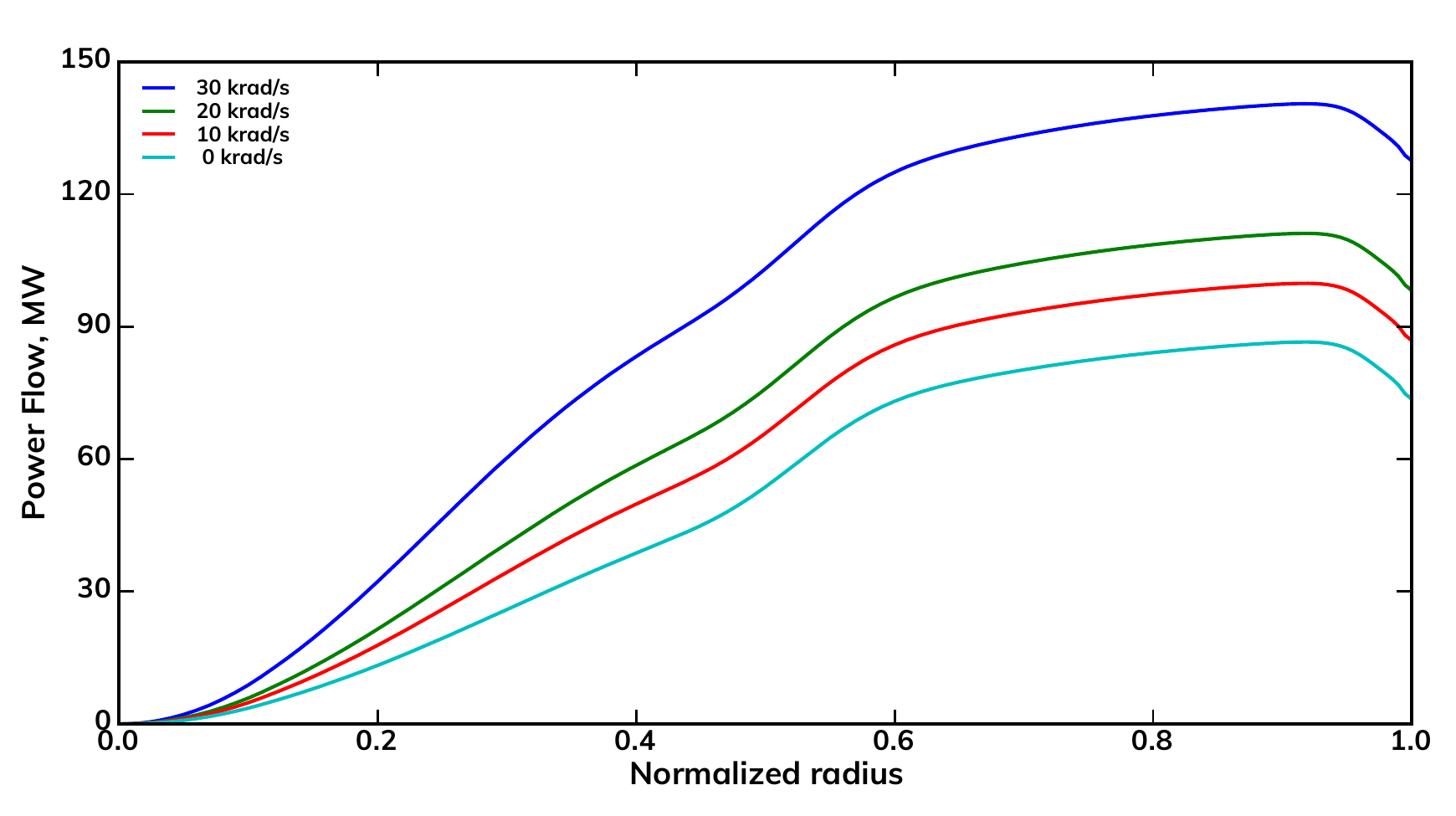}
\caption{Effect of intrinsic rotation on predicted power-flow profiles. Higher rotation weakly suppresses turbulence and increases $P_{\rm sep}$, but the total change remains $\lesssim 20\% $ over the range of amplitudes considered.}
\label{fig:RotationPowerFlow}
\end{figure}

The corresponding power‐flow profiles are shown in Fig.~\ref{fig:RotationPowerFlow}. Variations in rotation introduce a measurable but moderate change in $P_{\rm sep}$: the separatrix power increases with higher rotation amplitude, with a total change of $\lesssim 20\%$ across the full scan. This behavior reflects the reduction in ion heat conductivity when rotational shear stabilizes dominant turbulent modes in the mid‐radius region.

\begin{figure}[t]
\centering
\begin{minipage}[t]{0.48\textwidth}
  \centering
  \includegraphics[width=\linewidth]{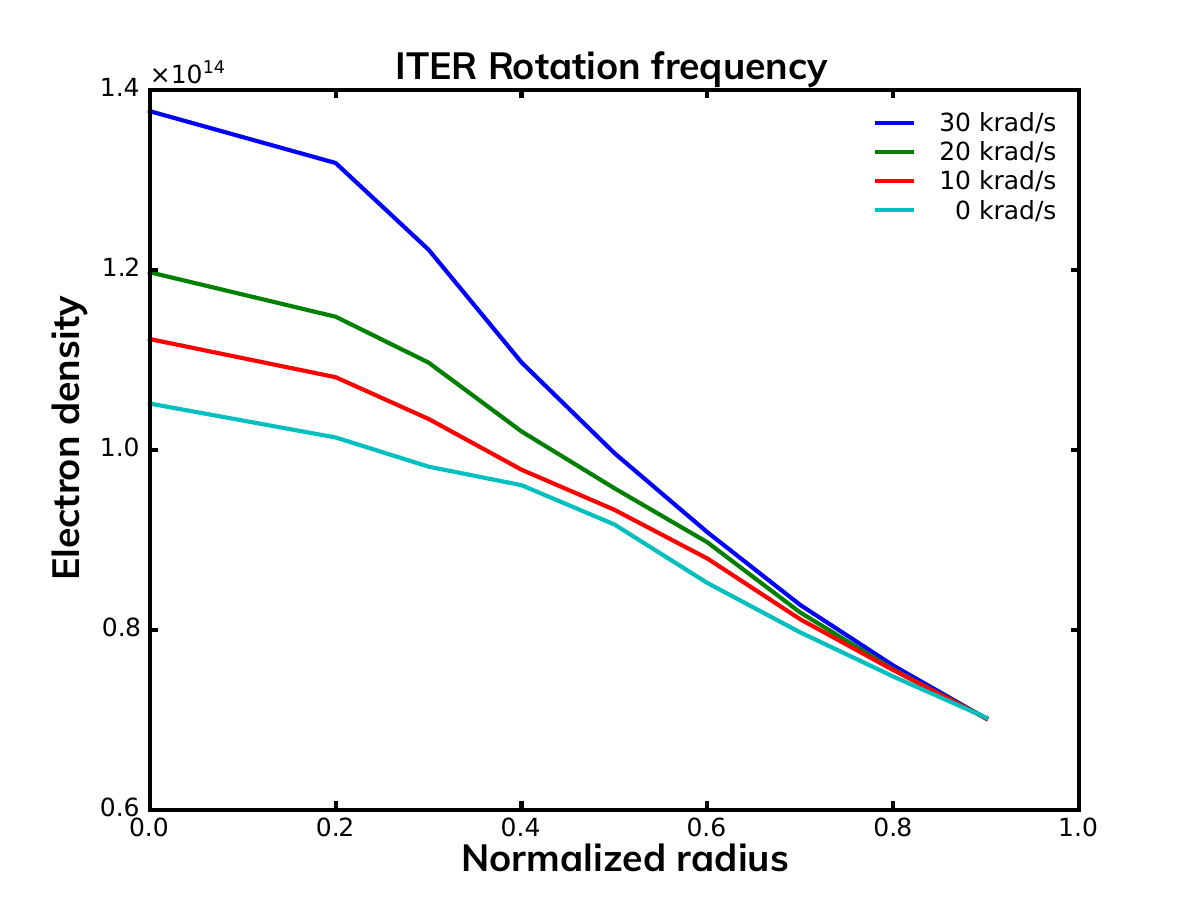}
\end{minipage}
\hfill
\begin{minipage}[t]{0.48\textwidth}
  \centering
  \includegraphics[width=\linewidth]{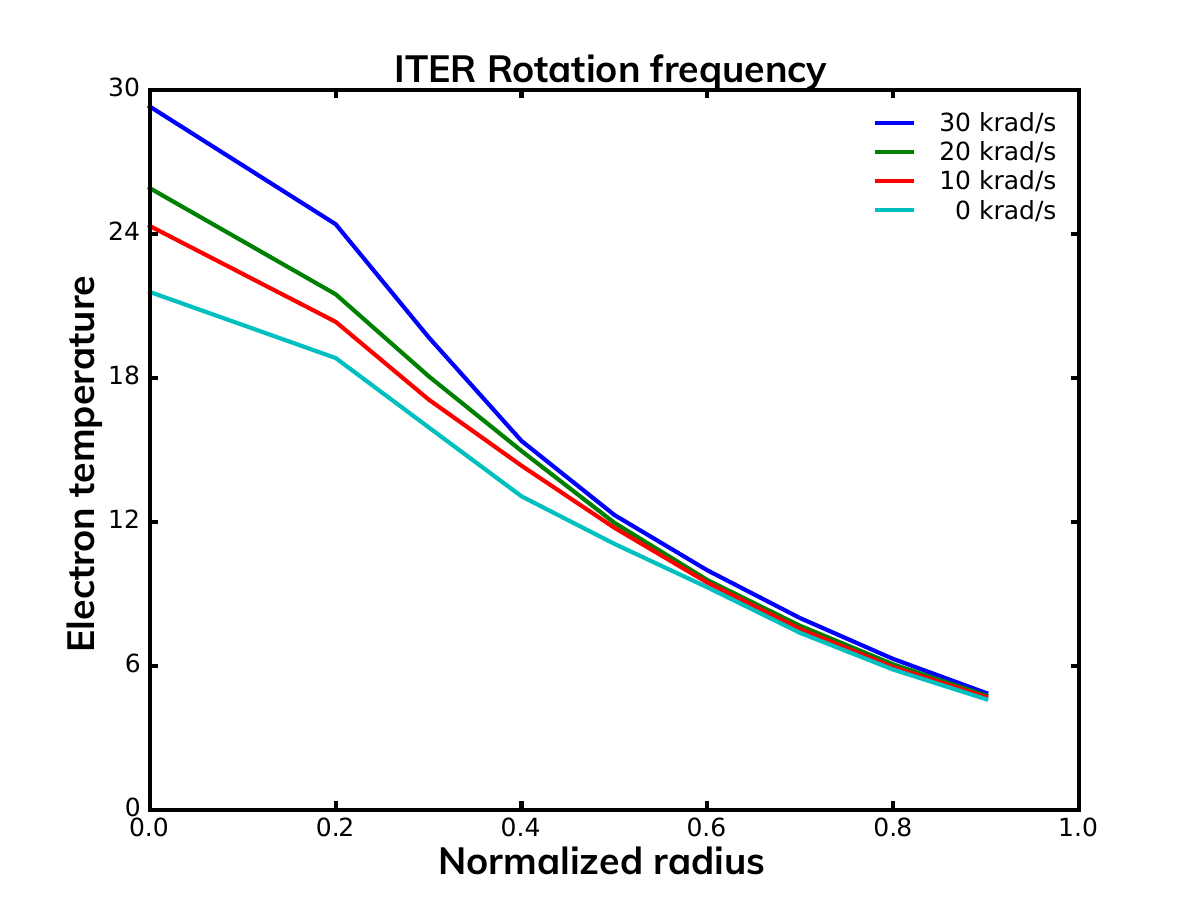}
\end{minipage}
\caption{Electron density (left) and electron temperature (right) profiles
predicted by TGYRO for varying prescribed toroidal rotation amplitudes in the
ITER baseline scenario. Increasing rotation weakly modifies the mid-radius
profiles through reduced turbulent transport, while pedestal values remain
largely unchanged.}
\label{fig:RotationProfiles_TeNe}
\end{figure}

The predicted changes in separatrix power shown in Fig.~\ref{fig:RotationPowerFlow} arise directly from the corresponding modifications to the electron temperature and density profiles. These profiles, computed self-consistently by TGYRO for each prescribed toroidal rotation amplitude, are shown in Fig.~\ref{fig:RotationProfiles_TeNe}. Increasing rotation leads to modest stiffening of the mid-radius profiles, consistent with reduced turbulent transport when rotational shear stabilizes dominant microinstabilities. As a result, both $T_e$ and $n_e$ increase slightly in the core and mid-radius, leading to an increase in the outward conductive heat flux and, consequently, in the power crossing the separatrix. Pedestal values remain largely unchanged, indicating that the rotation sensitivity of $P_{\rm sep}$ is governed primarily by core transport rather than edge boundary conditions.

These results indicate that while rotation can shift the predicted core power balance, the effect is significantly smaller than that arising from variations in $Z_{\rm eff}$ (Section~\ref{sec:PowerFlow}) or auxiliary heating (Section~\ref{sec:AuxPower}). Thus, uncertainties associated with intrinsic rotation do not alter the principal compatibility window identified in this study, provided that the rotation amplitude remains within the range expected from existing scaling projections.

%% file: Section7_Conclusions.tex
\section{Conclusions and outlook}
\label{sec:Conclusions}

This work identifies an ITER operational window consistent with simultaneously achieving $Q_{fus} \gtrsim 10$ and divertor-compatible exhaust levels at  —$Z_{\rm eff} \approx 1.6\text{--}1.75$ and $0.75 \lesssim f_{P_{\rm aux}} \le 1.0$—using SOLPS  power exhaust and TGYRO core performance predictions. We have performed an integrated, predict-first investigation of impurity concentration, auxiliary heating and intrinsic rotation impacts on core power-exhaust compatibility for the ITER 15~MA baseline scenario. Using the OMFIT STEP workflow, with TGYRO providing transport solutions calculated with the TGLF \cite{staebler_tglf_2007}  and NEO \cite{belli_neo_2008} transport models, we identify a narrow operating region in which fusion gain and divertor heat-flux requirements can be satisfied simultaneously.

A controlled $Z_{\rm eff}$ scan revealed that small variations in impurity content can shift the separatrix power flow by more than 50\%. Scenarios with $Z_{\rm eff} \simeq 1.75$ yield $P_{\rm sep} \simeq 100\,\mathrm{MW}$, in direct agreement with SOLPS-ITER modeling of no-ELM neon-seeded divertor solutions. Increasing $Z_{\rm eff}$ beyond this range further improves radiative power sharing in the scrape-off layer but dilutes the core plasma and reduces alpha heating, threatening the $\mathrm{Q} \ge 10$ ITER performance goals.

Additionally, we demonstrate that moderate reductions in external heating can produce similar power-flow outcomes for fixed pedestal conditions. For $Z_{\rm eff} = 1.6$, reducing auxiliary power to $f_{\mathrm{Paux}} \sim 0.75$ again results in $P_{\rm sep} \simeq 100\,\mathrm{MW}$. Rotation sensitivity studies confirmed that variations in toroidal flow magnitude over a reasonable range of low-torque operations translate into only modest variation ($\lesssim 20\%$) in the predicted power flows across the separatrix. Finally, detailed modeling of charge-exchange radiation using AURORA showed that neutrals inside the separatrix are predicted to be too rare to significantly alter impurity radiation or ionization balance in ITER core conditions.

Taken together, these results establish a physics-based compatibility window for ITER operation: $Z_{\rm eff} \approx 1.6$–$1.75$ and $0.75 \lesssim f_{\mathrm{Paux}} \le 1.0$, assuming metallic-wall conditions and neon seeding levels consistent with approved divertor protection targets. This provides actionable guidance for early ITER operational planning, including strategies for impurity-seeding control and auxiliary-heating scheduling.

Although we explored variations in rotation, the present work assumes either intrinsic or uniformly scaled rotation profiles rather than the low-but-finite $~P_{\mathrm{NBI}}$ torque expected in ITER. Even weak beam torque can generate modest shear at $\rho\sim0.5$–$0.7$, potentially affecting ion turbulent transport and impurity mixing \cite{Chrystal2017}. While the rotation sensitivity analysis presented here yielded a  $\lesssim 20\%$ variation in $P_{\mathrm{sep}}$, suggesting that the compatibility window is stable to reasonable flow variations, additional calculations quantifying the impacts of finite beam torque in these conditions would improve this initial assessment. Future work will therefore integrate full NBI torque deposition, momentum transport, and pedestal–core coupling to determine whether intrinsic rotation + weak beams can expand (or compress) the allowable $(Z_{\mathrm{eff}},P_{\mathrm{aux}})$ space.

Looking ahead, tighter coupling between SOLPS-ITER and STEP is needed to self-consistently evolve pedestal structure and scrape-off-layer conditions during transport relaxation. Quantifying uncertainties related to pedestal transport, recycling, and helium exhaust physics represents a key next step toward robust whole-device performance prediction. Further validation against high-power experiments on JET, ASDEX Upgrade and DIII-D will strengthen confidence in the predictive capability  demonstrated here.

The framework and findings reported in this work provide a foundation for integrated scenario optimization in ITER, highlighting the need for precise impurity and heating control to achieve performance targets while ensuring viable power exhaust. As ITER transitions from construction to operation, these predictive methodologies will play an essential role in realizing successful and sustained burning-plasma scenarios.